\documentclass[copyright]{eptcs}

\providecommand{\event}{GRAPHITE 2012}

\usepackage[pdftex]{graphicx}

\usepackage[all]{xy}
\UseComputerModernTips

\usepackage{float} 

\usepackage{amssymb}

\usepackage{listings} 

\newcommand{\nat}{\mathbb{N}}

\newcommand{\eventually}{\lozenge}
\newcommand{\nxt}{\bigcirc}

\title{A Comparison of Sequential and GPU Implementations of Iterative Methods to Compute Reachability Probabilities\thanks{This research is supported by the Natural Sciences and Engineering Research Council of Canada, an Ontario Graduate Scholarship and NVIDIA.}}
\author{Elise Cormie-Bowins
\institute{DisCoVeri Group, Department of Computer Science and Engineering, York University\\
4700 Keele Street, Toronto, ON, M3J 1P3, Canada}}
\def\titlerunning{A Comparison of Sequential and GPU Implementations}
\def\authorrunning{Elise Cormie-Bowins}

\begin{document}

\maketitle

\begin{abstract}
We consider the problem of computing reachability probabilities:
given a Markov chain, an initial state of the Markov chain, and a 
set of goal states of the Markov chain, what is the probability of
reaching any of the goal states from the initial state?  This problem
can be reduced to solving a linear equation
$\mathbf{A} \cdot \mathbf{x} = \mathbf{b}$ for \textbf{x}, where \textbf{A} 
is a matrix
and \textbf{b} is a vector.  We consider two iterative methods to
solve the linear equation: the Jacobi method and the 
biconjugate gradient stabilized (BiCGStab) method.  For both methods, a 
sequential and a parallel version have been implemented.  The
parallel versions have been implemented on the compute unified
device architecture (CUDA) so that they can
be run on a NVIDIA graphics processing unit (GPU).  From our experiments
we conclude that as the size of the matrix increases, the CUDA 
implementations outperform the sequential implementations.
Furthermore, the BiCGStab method performs better than the Jacobi 
method for dense matrices, whereas the Jacobi method does better 
for sparse ones.  Since the reachability probabilities problem
plays a key role in probabilistic model checking, we also compared
the implementations for matrices obtained from a probabilistic
model checker.  Our experiments support the conjecture by
Bosnacki et al.\ that the Jacobi method is superior to Krylov
subspace methods, a class to which the BiCGStab method belongs,
for probabilistic model checking.
\end{abstract}

\section{Introduction}

Given a Markov chain, an initial state of the Markov chain, and a 
set of goal states of the Markov chain, we are interested in the 
probability of reaching any of the goal states from the initial state.
This probability is known as the reachability probability.  These
reachability probabilities play a key role in several fields,
including probabilistic model checking (see, for example,
\cite[Chapter~10]{Katoen}) and performance evaluation (see, for example,
\cite{H98}).

 As we will sketch in Section~\ref{section:reachability},
computing reachability probabilities can be reduced to solving a
linear equation of the form 
$\mathbf{A} \cdot \mathbf{x} = \mathbf{b}$ for \textbf{x}, where \textbf{A} 
is an $n \times n$-matrix and \textbf{b} is an $n$-vector.
Although the equation can be solved by inverting the matrix \textbf{A},
such an approach becomes infeasible for large matrices due to the 
computational complexity of matrix inversion.  For instance, 
Gauss-Jordan elimination has time complexity $O(n^3)$ 
(see, for example, \cite[Chapter~2]{Strang}).  For large matrices, iterative methods
are used instead.

The iterative methods compute successive approximations to obtain a 
more accurate solution to the linear system at each iteration.  For
an overview of linear methods, we refer the reader to, for example, 
\cite[Chapter~2]{Templates}.   In 
this paper, we consider two linear methods, namely 
the Jacobi method and the biconjugate gradient 
stabilized (BiCGStab) method.  

We have implemented a sequential version and a parallel version of the
Jacobi and BiCGStab method.  The sequential versions have been
implemented in C.  The parallel versions have been implemented using NVIDIA's 
compute unified device architecture (CUDA).  It allows us to run
C code on a graphics processing unit (GPU) with hundreds of cores. Currently, CUDA is only supported by NVIDIA GPUs.

Our CUDA implementation of the Jacobi method is based on the
one described by Bosnacki et al.\ in \cite{Bosnacki,Bosnacki2}.
When we started this research, we were aware of the paper by 
Gaikwad and Toke \cite{GT10}
that mentions a CUDA implementation of the BiCGStab method.
Since we did not have access to their code, we implemented the
BiCGStab method in CUDA ourselves.

To compare the performance of our four implementations, we constructed
three sets of tests.  First of all, we randomly generated matrices
with varying sizes and densities.  Our experiments show that
the BiCGStab method is superior to the Jacobi method for denser
matrices.  They also demonstrate a consistent performance benefit from 
using CUDA to implement the Jacobi method.  However, we observe that
the CUDA version of the BiCGStab method is only beneficial for larger, 
denser matrices.  For the smaller and sparser matrices, the sequential 
version of the BiCGStab method outperforms the CUDA version.

Secondly, we used an extension of the model checker Java PathFinder
(JPF) to generate matrices.  JPF can check properties of Java code, such as the
absence of uncaught exceptions.  Zhang \cite{Xin}
created a probabilistic extension of JPF that can
check properties of randomized sequential algorithms.
We used this extension to generate  transition probability matrices 
corresponding to the Java code of two randomized sequential algorithms.
The Jacobi method performed better than the BiCGStab method for
these matrices.  This supports the conjecture by
Bosnacki et al.\ \cite{Bosnacki,Bosnacki2} that the Jacobi method 
is superior to Krylov subspace methods, a class to which the 
BiCGStab method belongs, for probabilistic model checking.

Finally, we randomly generated matrices with the same sizes and
densities as the matrices produced by JPF.  We obtained very similar
results.  This suggests that size and density are the main 
determinants of which implementation performs best on probabilistic model 
checking data, and whether CUDA will be beneficial, rather than other 
properties unique to matrices found in probabilistic model checking.

\section{Reachability Probabilities of a Markov Chain}
\label{section:reachability}

In this paper, the term ``Markov chain" refers to a discrete-time Markov chain. Below, we review the well-known problem of computing the reachability
probabilities of a Markov chain. Our presentation is based on
\cite[Section~10.1]{Katoen}.  A \emph{Markov chain} 
is a triple $\mathcal{M}  = (S, \mathbf{P}, s_0)$ consisting of
\begin{itemize}
\item 
a nonempty and finite set $S$ of \emph{states}, and
\item 
a \emph{transition probability function} 
$\mathbf{P} : S \times S \to [0, 1]$ satisfying for all $s \in S$,
\begin{displaymath}
\sum_{s' \in S} \mathbf{P}(s, s') = 1,
\end{displaymath}
\item and an \emph{initial state} $s_0 \in S$.
\end{itemize}
The transition probability function can be represented as a matrix.
This matrix has rows and columns corresponding to the states of the 
Markov chain, and entries representing the probabilities of the 
transitions between these states.  For example, the probability 
transition function of the Markov chain depicted by
\begin{displaymath}
\xymatrix{
&& *=<20pt>[o][F]{0} \ar[dr]^{0.5} \ar@/_/[dl]_{0.5}\\
*=<20pt>[o][F]{1} \ar@(l,d)[]_{1} & *=<20pt>[o][F]{2} \ar[l]_{0.6} \ar@/_/[ur]_{0.4} && *=<20pt>[o][F]{3} \ar@(r,d)[]^{1}
}
\end{displaymath}
can be represented by the matrix
\begin{displaymath}
\left[
\begin{array}{cccc}
0 & 0 & 0.5 & 0.5\\
0 & 1 & 0 & 0\\
0.4 & 0.6 & 0 & 0 \\	
0 & 0 & 0 & 1
\end{array} 
\right]
\end{displaymath}

A \emph{path} of a Markov chain $\mathcal{M}$ is an infinite sequence
$s_1 s_2 \ldots$ of states such that $\mathbf{P}(s_i, s_{i+1}) > 0$ 
for all $i \geq 1$.  We denote the set of paths of $\mathcal{M}$ that
start in state~$s$ by $\mathit{Paths}(s)$.  For the above 
Markov chain, the set of paths starting in state $0$ is 
$\{\, (02)^n 1^{\omega} \mid n \in \nat \,\} \cup
\{\, (02)^n $ $ 0 $ $3^{\omega} \mid n \in \nat \,\} \cup \{ (02)^{\omega} \}$.

To define the probability of events such as eventually reaching a 
set of states from the initial state, we associate a probability
space with a Markov chain $\mathcal{M}$ and a state $s$.  Recall that 
a probability space consists of a set, a $\sigma$-algebra, and a probability
measure.  In this case, the set is $\mathit{Paths}(s)$.
The $\sigma$-algebra is generated by the so-called cylinder sets.
Given a finite sequence of states $s_1 \ldots s_n$, its \emph{cylinder set}
$\mathit{Cyl}(s_1 \ldots s_n)$ is defined by
\begin{displaymath}
\mathit{Cyl}(s_1 \ldots s_n)
=
\{\, \pi \in \mathit{Paths}(s) \mid s_1 \ldots s_n
\mbox{ is a prefix of } \pi \,\}.
\end{displaymath}
For the above Markov chain, we have that
$\mathit{Cyl}(03) = \{ 0 $ $ 3^{\omega} \}$.  We define
\begin{displaymath}
\mathit{Pr}(\mathit{Cyl}(s_1 \ldots s_n))
=
\prod_{1 \leq i < n} \mathbf{P}(s_i, s_{i+1}).
\end{displaymath}
Recall that \textit{Pr} can be uniquely extended to a probability
measure on the $\sigma$-algebra generated by the cylinder sets.
For the above Markov chain, we have that
$\mathit{Pr}(\mathit{Cyl}(03)) = 0.5$.

\subsection{Property of Interest}

In the following, we are interested in two particular events.
First of all, given a Markov chain $\mathcal{M}$, a state $s$
of the Markov chain and a set of states \textit{GS}, known
as the \emph{goal states}, of the Markov chain, we are interested in 
the probability of reaching a state in \textit{GS} in one
transition when starting in $s$.  We denote the set of paths
starting in $s$ that reach a state in \textit{GS} in one transition
by $\{\, \pi \in \mathit{Paths}(s) \mid \pi \models \nxt \mathit{GS} \,\}$.
This set can be shown to be measurable, that is, it belongs to
the $\sigma$-algebra.  Its probability we denote by
$\mathit{Pr}(s \models \nxt \mathit{GS})$.

Secondly, given a Markov chain $\mathcal{M}$, a state $s$ of the 
Markov chain and a set of goal states \textit{GS}, we are interested 
in the probability of reaching a state in \textit{GS} in zero or
more transitions when starting in $s$.  We denote the set of paths
starting in $s$ that reach a state in \textit{GS} in zero or more transitions
by $\{\, \pi \in \mathit{Paths}(s) \mid \pi \models \eventually \mathit{GS} \,\}$.
Also this set can be shown to be measurable.  Its probability we denote by
$\mathit{Pr}(s \models \eventually \mathit{GS})$.

The problem we are examining in this paper is the following: Given a 
Markov chain $\mathcal{M}$, and set of goal states \textit{GS}, what is 
the probability to reach a state in \textit{GS} in zero or more transitions
from the initial state?  In other words, what is the probability of states 
in \textit{GS} eventually being reached?  That is, we want to compute
$\mathit{Pr}(s_0 \models \eventually \mathit{GS})$, where $s_0$ is
the initial state of $\mathcal{M}$.  This is what is referred to as 
computing \emph{reachability probabilities} in \cite[Chapter~10.1.1]{Katoen}.

\subsection{Partitioning the Set of States}

To compute the reachability probabilities, one usually first partitions
the set of states of the Markov chain into three parts, based on their
probability to reach the set \textit{GS} of goal states:
\begin{itemize}
\item 
$S_{=1} = \{\, s \in S \mid \mathit{Pr}(s \models \eventually \mathit{GS}) = 1 \,\}$ 
\item 
$S_{=0} = \{\, s \in S \mid \mathit{Pr}(s \models \eventually \mathit{GS}) = 0 \,\}$
\item 
$S_? = S \setminus (S_{=1} \cup S_{=0})$
\end{itemize}
The partitioning of the set of states can be done easily by considering 
the \emph{underlying digraph} of the Markov chain.  The vertices of this 
digraph are the states of the Markov chain.  There is an edge from 
state~$s$ to state~$s'$ if and only if $\mathbf{P}(s, s') > 0$. 
Using graph algorithms, such as depth-first-search or breadth-first-search,
the set of states can be partitioned as follows:
\begin{itemize}
\item 
To find $S_{=0}$, determine the set of all states that can reach 
\textit{GS} (including the states in \textit{GS}). The complement of this 
set is $S_{=0}$.
\item 
To find $S_{=1}$, determine the set of all states that can reach $S_{=0}$. 
The complement of this set is $S_{=1}$. 
\item 
To find $S_?$, simply take the complement of $S_{=1} \cup S_{=0}$.
\end{itemize}
Consider the above Markov chain and let $3$ be the only goal state.
Then $S_{=0} = \{ 1 \}$, $S_{=1} = \{ 3 \}$ and $S_? = \{ 0, 2 \}$.

\subsection{Computing Reachability Probabilities by State}
\label{computing_reachability}

To compute the reachability probabilities, one must determine the 
probability of the initial state leading to a state in \textit{GS},
that is, one has to compute $\mathit{Pr}(s_0 \models \eventually \mathit{GS})$.
To do so, one must also determine the probabilities of reaching a state
in \textit{GS} from other states.

For each state $s$ we compute $x_s$, which is the probability of 
reaching \textit{GS} from $s$, that is, 
\linebreak
$x_s = \mathit{Pr}(s \models \eventually \mathit{GS})$.  The values of 
$x_s$, for $s \in S$, can be expressed as a vector, $\mathbf{x}$.
For any state $s \in S_{=1}$, by definition $x_s = 1$. 
Similarly, for each $s \in S_{=0}$, $x_s = 0$. So once the states have 
been partitioned into the three sets described above, the only values of  
$\mathbf{x}$ that need to be calculated are $\{\, x_s \mid s \in S_? \,\}$.
These values satisfy the following equation:
\begin{displaymath}
x_s = \sum_{s' \in S \setminus \mathit{GS}} \mathbf{P}(s, s') \cdot x_{s'} 
+ \sum_{s' \in \mathit{GS}} \mathbf{P}(s, s').
\end{displaymath}
So, we will create a matrix $\mathbf{A}$, which includes only the transition 
probabilities between states in $S_?$.  For each $s$, $s' \in S_?$,
$\mathbf{A}_{s, s'} = \mathbf{P}(s, s')$.  To aid in calculations, we will 
also create a vector $\mathbf{b}$.  For each
$s \in S_?$, $b_s = \mathit{Pr}(s \models \nxt \mathit{GS})$, that
is, the probability of a state in \textit{GS} being reached from $s$ 
in one transition.  Consider the above Markov chain and let $3$ be
the only goal state.  Then states 1 and 3 are excluded because they do not belong to $S_?$, and
\begin{displaymath}
\mathbf{A} =
\left [
\begin{array}{cc}
0 & 0.5\\
0.4 & 0
\end{array}
\right ]
\mbox{ and }
\mathbf{b} = 
\left [
\begin{array}{cc}
0.5\\
0
\end{array}
\right ]
\end{displaymath}

The equation for $\mathbf{x}$ above can be written as 
$\mathbf{x} = \mathbf{A} \cdot \mathbf{x} + \mathbf{b}$.  Rearranged, 
this becomes $(\mathbf{I} - \mathbf{A}) \cdot \mathbf{x} = \mathbf{b}$, where
$\mathbf{I}$ is an identity matrix.
$\mathbf{A}$ and $\mathbf{b}$ are already known from the Markov chain.
So, $\mathbf{x}$ can be found by solving the linear equation 
$(\mathbf{I} - \mathbf{A}) \cdot \mathbf{x} = \mathbf{b}$ for 
$\mathbf{x}$.

There are several ways to solve the equation 
$(\mathbf{I} - \mathbf{A}) \cdot \mathbf{x} = \mathbf{b}$ for $\mathbf{x}$. 
Most obviously, one can find 
\linebreak
$(\mathbf{I} - \mathbf{A})^{-1}$. However, 
methods to find matrix inverses tend to have high computational complexity 
and, hence, for large matrices this becomes infeasible. For instance, 
Gauss-Jordan elimination has time complexity $O(n^3)$ 
(see, for example, \cite[Chapter~2]{Strang}).

Iterative approximation methods find solutions that are within a specified 
margin of error of the exact solutions, and can work much more quickly. 
The methods used in this paper are in this category, and will be discussed 
in more detail next.

\section{Iterative Methods}
\label{iterative}

Solving the linear equation $(\mathbf{I} - \mathbf{A}) \cdot \mathbf{x} = \mathbf{b}$ for $\mathbf{x}$ 
can be very time-consuming, especially when \textbf{A} is large. To solve this equation, there are numerous iterative methods.  These methods compute successive 
approximations to obtain a more accurate solution to the linear system at each iteration.  For
an overview of linear methods, we refer the reader to, for example, \cite[Chapter~2]{Templates}.

These iterative methods can be classified into two groups: the stationary methods and the 
nonstationary ones.  In stationary methods, the same information
is used in each iteration.  As a consequence, these methods are usually easier to understand and implement.
However, convergence of these methods may be slow.  In this paper, we consider one
stationary linear method, namely the Jacobi method.

In nonstationary methods, the information used may change per iteration.  These methods
are usually more intricate and harder to implement, but often give rise to faster
convergence.  In this paper we focus on one particular nonstationary linear method,
namely the biconjugate gradient stabilized (BiCGStab) method.  

The BiCGStab method was developed by Van der Vorst \cite{Vorst92}.  It is a type of Krylov 
subspace method.  Unlike most of these methods, it can solve non-symmetric linear systems, 
and was designed to minimize the effect of rounding errors.  If exact arithmetic 
is used, it will terminate in at most $n$ iterations for an $n \times n$ matrix 
\cite[page~636]{Vorst92}.  In practice, it often requires much fewer iterations to find 
an approximate solution.

The un-preconditioned version of this method was used, as the sequential and GPU performance 
of the preconditioning method would be a separate issue.  For some matrices a preconditioner 
is necessary to use the BiCGStab method, but that is not the case for the data encountered here. 
Buchholz \cite{Buchholz} did a limited comparison between a preconditioned and un-preconditioned version of the BiCGStab method specifically on matrices that represent stochastic processes, and it does 
not show a large performance difference.

\subsection{The Jacobi Method}
\label{the_jacobi_method}

Given the matrix \textbf{A} and the vector \textbf{b}, the Jacobi
method returns an approximate solution for \textbf{x} in the 
linear equation $\mathbf{A} \cdot \mathbf{x} = \mathbf{b}$.
Below, we only present its pseudocode.  For a detailed discussion
of the Jacobi method, we refer the reader to, for example,
\cite[Section~2.2.1]{Templates}.

\lstdefinelanguage{pseudo}{morekeywords={repeat,until,return,for,if,then,all,do}}
\begin{lstlisting}[language=pseudo,numbers=left,numberstyle=\tiny,mathescape=true]
Jacobi($\mathbf{A}, \mathbf{b}$):
$\mathbf{x}$ := random vector
repeat
   $\mathbf{x}'$ := $\mathbf{x}$
   for all $i = 1 \ldots n$ do
      $x_i$ := $\frac{1}{A_{i, i}} \cdot (b_i - \sum_{j \not= i} A_{i, j} \cdot x'_j)$
until $\mathbf{x}$ is accurate enough
return $\mathbf{x}$
\end{lstlisting}

\subsection{The BiCGStab Method}
\label{sect_bic}

Like the Jacobi method, the BiCGStab method returns an approximate solution 
to $\mathbf{A} \cdot \mathbf{x} = \mathbf{b}$ for a given matrix 
\textbf{A} and vector \textbf{b}.
Again, we only present the pseudocode.  For more details, we refer the
reader to the highly cited paper \cite{Vorst92} in which Van der Vorst 
introduced the method.

\begin{lstlisting}[language=pseudo,numbers=left,numberstyle=\tiny,mathescape=true]
BiCGStab($\mathbf{A}, \mathbf{b}$):
$\mathbf{x}$ := random vector
$\mathbf{r}$ := $\mathbf{b} - \mathbf{A} \cdot \mathbf{x}$
$\mathbf{q}$ := random vector such that $\mathbf{q} \cdot \mathbf{r} \neq 0$
$y$ := 1; $a$ := 1; $w$ := 1; $\mathbf{v}$ := $\mathbf{0}$; $\mathbf{p}$ := $\mathbf{0}$
repeat
   $y'$ := $y$
   $y$ := $\mathbf{q} \cdot \mathbf{r}$
   $\mathbf{p}$ := $\mathbf{r} + \frac{y \cdot a}{y' \cdot w} \cdot (\mathbf{p} - w \cdot \mathbf{v})$
   $\textbf{v}$ := $\textbf{A} \cdot \textbf{p}$
   $a$ := $\frac{y}{\mathbf{q} \cdot \mathbf{v}}$
   $\mathbf{s}$ := $\mathbf{r} - a \cdot \mathbf{v}$
   $\mathbf{t}$ := $\mathbf{A} \cdot \mathbf{s}$ 
   $w$ := $\frac{\mathbf{t} \cdot \mathbf{s}}{\mathbf{t} \cdot \mathbf{t}}$
   $\mathbf{x}$ := $\mathbf{x} +  a \cdot \mathbf{p}  +  w \cdot \mathbf{s}$
   $\mathbf{r}$ := $\mathbf{s} - w \cdot \mathbf{t}$
until $\mathbf{x}$ is accurate enough
return $\mathbf{x}$
\end{lstlisting}

\section{General Purpose Graphics Processing Units (GPGPU)}

Graphics processing units (GPUs) were primarily developed to improve 
the performance of graphics-intensive programs.  The market driving 
their production has traditionally been video game players. They are 
a throughput-oriented technology, optimized for data-intensive calculations 
in which many identical operations can be done in parallel on different 
data.  Unlike most commercially available multi-core central processing units (CPUs), which normally 
run up to eight threads in parallel, GPUs are designed to run hundreds of 
threads in parallel.  They have many more cores than CPUs, but these cores 
are generally less powerful.  GPUs sacrifice latency and single-thread 
performance to achieve higher data throughput \cite{Garland2010}.

Originally, GPU APIs were designed exclusively for graphics use, and 
their underlying parallel capabilities were difficult to access for 
other types of computation.  This changed in 2006 when NVIDIA released 
CUDA, the first architecture and API designed to allow GPUs to be used 
for a wider variety of applications \cite[Chapter~1]{Kirk2010}.  

All parallel algorithms discussed in this paper were implemented using CUDA. Different generations
of CUDA-enabled graphics cards have different compute capability levels, 
which indicate the sets of features that they support.  The graphics card 
used here is the NVIDIA GTX260, which has compute capability~1.3.  Cards 
with higher compute capability have been released by NVIDIA since the 
GTX260 was developed.  

The GTX260 has 896MB global on-card memory.  This GPU supports 
double-precision floating-point operations, which is required for BiCGStab. 
Lower compute capabilities (1.2 and below) do not support this.  The
GTX260 does not support atomic floating-point operations.  This limits the 
way operations that require dot-products and other row sums can be separated 
into threads.  Since these sums require all elements of a row to be summed 
to a global variable, and atomic addition can not be used to protect the 
integrity of the sum, the structure of the threads must do so. Here, we 
have simply created one thread per row and done these additions sequentially. 
The next generation of NVIDIA GPUs (compute capability 2.0 and higher) 
support atomic floating-point addition, though only on single-precision 
numbers.

Current GPUs have limited memory to store matrix data.  For all but the 
smallest matrices, some sort of compression is necessary so that they can 
be transferred to the GPU in their entirety.

For this project, data is stored using a compressed row storage, as 
described in \cite[page~57]{Templates}.  In this format, only the 
non-zero elements of the matrix are recorded.  The size of an
$n \times n$ matrix with $m$ non-zero elements compressed in this manner 
is $O(n + m)$, whereas uncompressed it is $O(n^2)$.  This representation 
saves significant space when used to store sparse matrices.

A compressed row matrix consists of three vectors.  
The vector \textit{rstart} contains integers and has length~$n + 1$, where $\mathit{rstart}_i$ is the total number of non-zero elements in the 
first $i$ rows.
The vector \textit{col} also contains integers and has length~$m$. It stores the column position of each non-zero element.
The vector \textit{nonzero} contains floating points and has
length~$m$.  It stores the non-zero elements of the matrix.

For example, the matrix
\begin{displaymath}
\left [
\begin{array}{ccc}
1 & 0 & 0\\
0 & 0 & 2\\
0 & 0 & 3
\end{array} 
\right ]
\end{displaymath}
is represented by the vectors
\begin{center}
\begin{tabular}{rcl}
\textit{rstart} & : & [0, 1, 2, 3]\\
\textit{col} & : & [0, 2, 2]\\
\textit{nonzero} & : & [1, 2, 3]
\end{tabular}
\end{center}

\section{Parallel Implementations in CUDA}

Next, we present parallel versions of the Jacobi method and the
BiCGStab method.  These parallel versions are implemented in CUDA.

\subsection{The Jacobi Method}

The Jacobi method was parallelized as in \cite{Bosnacki}.  Given an $n \times n$-matrix \textbf{A} 
and an $n$-vector \textbf{b}, in the parallel implementation of the Jacobi method $n$ threads are 
created.  For each thread $i$, where $0 \leq i < n$, the following algorithm is used.  Essentially, one thread traverses each row of the 
matrix \textbf{A} to compute the corresponding element of \textbf{x}. Thus, each element of \textbf{x} is computed in paralllel.


\begin{lstlisting}[language=pseudo,numbers=left,numberstyle=\tiny,mathescape=true]
$i$ := thread id
if $i$ = 0 then
   $\mathit{terminate}$ := true
if $i < n$ then
   $d$ := $b_i$
   $l$ := $\mathit{rstart}_i$
   $h$ := $\mathit{rstart}_{i+1} - 1$
   for all $j = l \ldots h$ do
      $d$ := $d - \mathit{nonzero}_j \cdot x_{col_j}$
   $x_i$ := $\frac{d}{A_{i,i}}$
   if $x_i$ is not accurate enough then
     $\mathit{terminate}$ = false
\end{lstlisting}
The variable \textit{terminate} is shared by all the threads to
determine when the iteration can stop.  Note that line~5--10 of the 
code above corresponds to line~5--6 of the Jacobi method in
Section~\ref{the_jacobi_method}.  Thus, this algorithm represents one iteration of the Jacobi method. The main program repeatedly launches the corresponding code on the GPU, waiting between executions for all threads to complete, until \textbf{x} reaches the desired accuracy.

\subsection{The BiCGStab Method}

Each iteration of the BiCGStab  method consists of several matrix and vector multiplications, that each result in a vector. As most of these must be done in order, they were parallelized individually by creating one thread for each element of the resulting vectors.

For instance, the operation in line 15 is split between $n$ threads, so for each $0 \leq i < n$ a thread does the following:
\begin{displaymath}
x_i = x_i + a \cdot p_i + w \cdot s_i
\end{displaymath}
And thus each element of the vector is calculated in parallel. The matrix operation in line 10 is done as follows for each of the $n$ threads:
\begin{displaymath}
v_i =  \mathbf{A}_i \cdot \mathbf{p}
\end{displaymath}
where $\mathbf{A}_i$ denotes the $i^{\mathrm{th}}$ row of the matrix
\textbf{A}.
Dot products are done sequentially.  It would be possible to increase parallelism by splitting these into multiple sums done in parallel. This may require too much overhead to result in a significant performance gain.

Currently, CUDA requires all threads running on a GPU to execute the same code. There are some separate steps of the BiCGStab method which could be executed in parallel, but this is not possible with a single GPU of the type used here. This may be possible using the next generation of GPUs, or multiple GPUs. However, most steps of the algorithm must be done in sequence.

Below is an abbreviated version of the CUDA code used to implement BiCGStab. To save space, non-essential code has been removed, and some steps are summarized in square brackets. Kernel numbering corresponds to the line numbers of the algorithm in Section \ref{sect_bic}. Kernels for subsequent steps are combined when they require the same number of threads. Sequential steps are done on the GPU with a single thread, since this avoids time-consuming data transfers between the GPU and host computer.

The first portion of the code, below, executes on the CPU.  Pointers prefixed by $d\_$ indicate data stored on the GPU, and $n$ is the dimension of the matrix.  The matrix \textbf{A} is represented in row-compressed form on the GPU as \emph{d\_col, d\_rstart} and \emph{d\_nonzero}. 


\footnotesize 

\begin{verbatim}
int terminate = 0;
for(int i = 0; i <= max_iterations && !terminate; i++){   		
     [d_yprime = d_y, switch pointers without moving data]
     step8Kernel<<<1,1>>>(d_B, d_y, d_yprime, d_a, d_w, d_q, d_r, n);
     step9Kernel<<<grid,block>>>(d_p, d_r, d_B, d_w, d_v, n, blocksz);
     matrixVectorMult<<<grid,block>>>(d_col, d_rstart, d_nonzero, d_p, d_v, n, blocksz);
     step11Kernel<<<1,1>>>(d_a, d_y, d_q, d_v, n);
     step12Kernel<<<grid,block>>>(d_s, d_r, d_a, d_v, n, blocksz);
     matrixVectorMult<<<grid,block>>>(d_col, d_rstart, d_nonzero, d_s, d_t, n, blocksz);
     step14Kernel<<<1,1>>>(d_w, d_t, d_s, n);
     step15_16Kernel<<<grid,block>>>(d_x, d_a, d_p, d_w, d_s, d_r, d_t, n, blocksz, d_terminate);
     [terminate = d_terminate, transferred from GPU to host machine]
}
\end{verbatim}

\normalsize

The kernel methods below execute on the GPU. The kernels for steps 12 are 14 are very similar to previous steps, and are excluded. The \emph{matrixVectorMult} method, not shown, is a generic parallel matrix-vector multiplication kernel.

\vspace{6pt}
\footnotesize 

\begin{verbatim}
__global__ static void step8Kernel(double *d_B, double *d_y, double *d_yprime, double *d_a, 
double *d_w, double *d_q, double *d_r, int n){
     double result = 0;
     for(int i = 0; i < n; i++){ result += d_q[i] * d_r[i]; }
     *d_y = result;	
     *d_B = (*d_y / *d_yprime) * (*d_a / *d_w); //prepare scalar for step 9
}

__global__ static void step9Kernel(double *d_p, double *d_r, double *d_B, double *d_w, 
double *d_v, int n, int blocksz){
     int i = blockIdx.x * blocksz + threadIdx.x; //thread index
     if(i < n){ d_p[i] = d_r[i] + *d_B * ( d_p[i] - *d_w * d_v[i]); }
}

__global__ static void step11Kernel(double *d_a, double *d_y, double *d_q, double *d_v, int n){	
     double dot_q_v = 0; //dot product of q,v
     for(int i = 0; i < n; i++){ dot_q_v += d_q[i] * d_v[i]; }
     *d_a = *d_y / dot_q_v;
}

__global__ static void step15_16Kernel(double *d_x, double *d_a, double *d_p, double *d_w, 
double *d_s, double *d_r, double *d_t, int n, int blocksz, int* d_Terminate){
     int i = blockIdx.x * blocksz + threadIdx.x; //thread index
     if(i == 0){ *d_Terminate = 1; } //one thread sets d_Terminate
     if(i < n){		
          d_x[i] = d_x[i] + *d_a * d_p[i] + *d_w * d_s[i]; //step 15		
          double diff = abs(d_s[i]); 	
          if(diff > ERROR){ *d_Terminate = 0;} //stop when residual is near zero
          d_r[i] = d_s[i] - *d_w * d_t[i]; //step 16
     }
}
\end{verbatim}

\normalsize

\section{Performance on Randomly Generated Matrices}
\label{perf_random}

For these tests, random matrices were generated. The entries were random 
positive integers, placed in random locations.  The matrices were then 
modified to have non-zero diagonal entries and be diagonally-dominant, which 
ensured that both the Jacobi and BiCGStab methods were able to solve 
equations containing them.  For each matrix \textbf{A}, a vector \textbf{b}
of random integers was also created. This formed the equation 
$\mathbf{A} \cdot \mathbf{x} = \mathbf{b}$, to be solved for \textbf{x}. 
Each trial used a newly-generated matrix and vector.

\begin{figure}[!ht]
\centering
 \fbox{ 
\includegraphics[width=\linewidth]{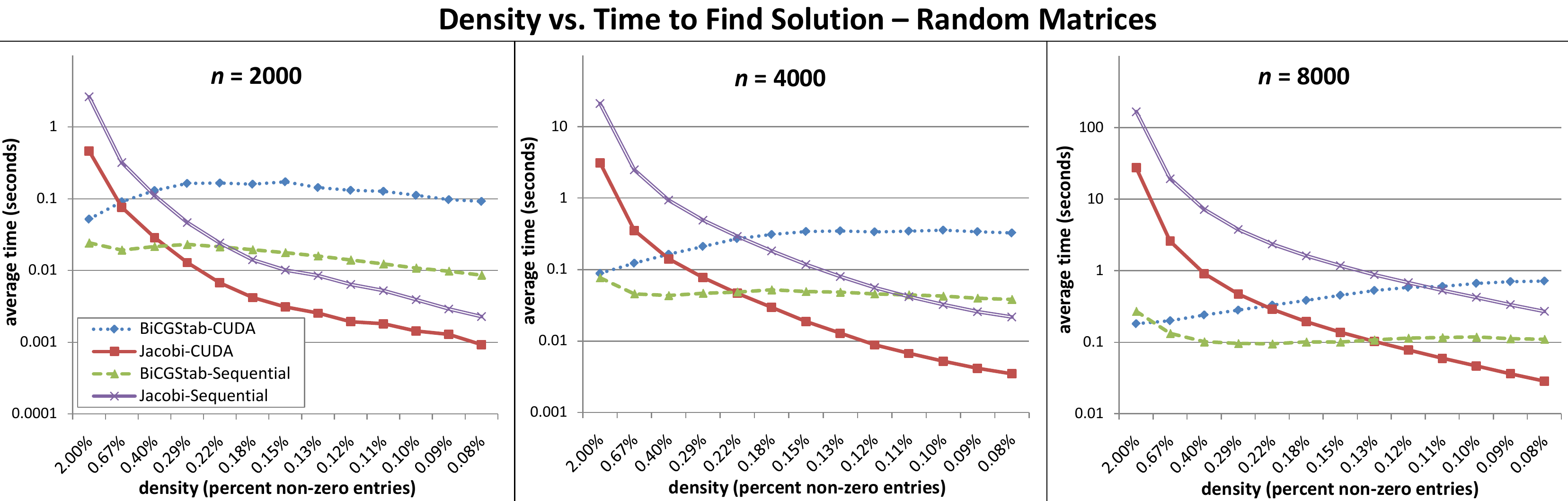}
}
\caption{The effect of varying density on performance, for three matrix sizes. A logarithmic scale is used on the y-axis. Average times are based on 20 trials. The standard deviations of the times measured are too small to be shown.}
\label{DensityVsTime}
\end{figure}

Figure~\ref{DensityVsTime} shows the performances of the four implementations 
on random matrices of several sizes, and varying densities.  The matrix 
densities are similar to those encountered in probabilistic model checking.  
These graphs indicate that the implementations' relative performances change 
their order as the density increases.  Furthermore, as the size of the matrix 
increases, the density at which the performance order changes becomes lower. 
Thus, which implementation performs best depends on the size and density of 
the matrix it is being used on.

Generally, the graphs in Figure~\ref{DensityVsTime} show that the BiCGStab 
method is superior to the Jacobi method for denser matrices, but the Jacobi method performs best for very sparse matrices.  They also 
demonstrate a consistent performance benefit from using CUDA to implement 
the Jacobi method.  However, the third graph shows that the CUDA version 
of the BiCGStab method is only beneficial for larger, denser matrices.  For other matrices, the sequential version of the BiCGStab 
method outperforms the CUDA version.

\begin{figure}[!ht]
\centering
 \fbox{ 
\includegraphics[width=0.7\linewidth]{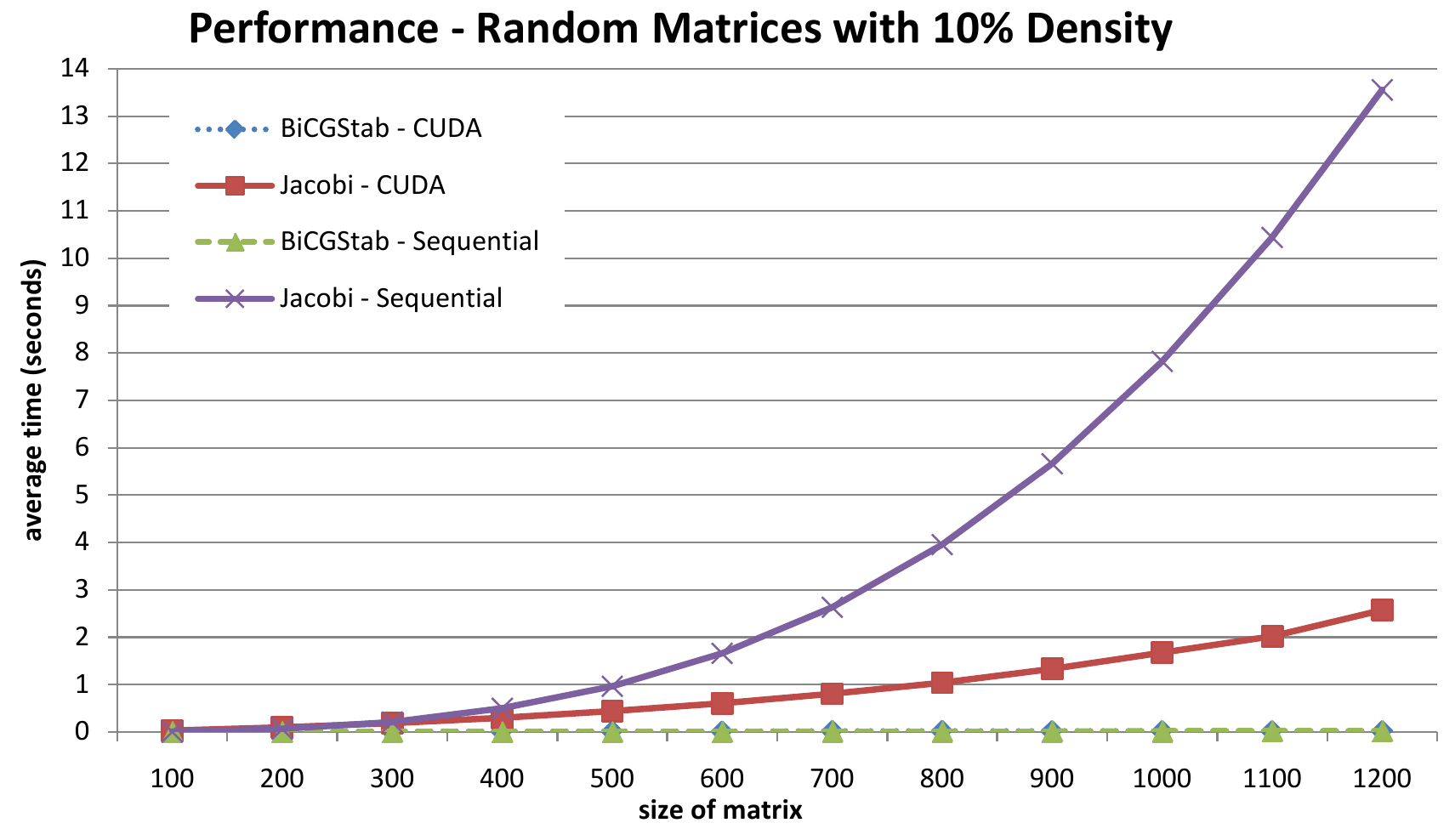}
}
\caption{Performance on randomized matrices of varying sizes, all with 10\% density.}
\label{Random_density10_all}
\end{figure}

\begin{figure}[!ht]
\centering
 \fbox{ 
\includegraphics[width=0.7\linewidth]{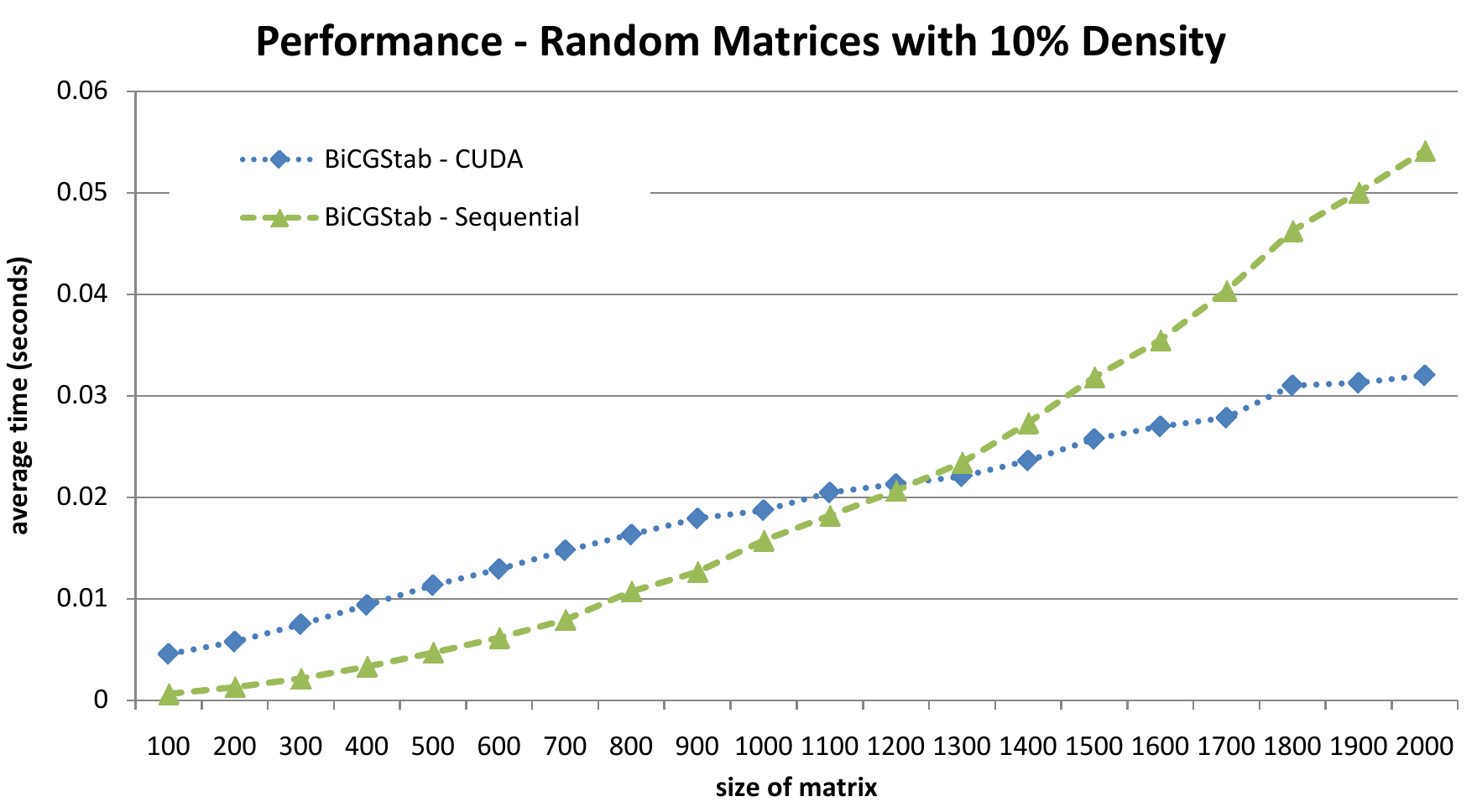}
}
\caption{BiCGStab performance on randomized matrices of varying sizes, all with 10\% density.}
\label{Random_density10_BiC}
\end{figure}

To confirm this, Figure~\ref{Random_density10_all} and 
Figure~\ref{Random_density10_BiC} show the implementations' performances on random matrices 
with 10\% density.  As the sizes of these relatively dense matrices 
increase, the CUDA versions of both implementations increasingly outperform 
their sequential counterparts, and the BiCGStab method significantly 
outperforms the Jacobi method.

\section{Performance on Probabilistic Model Checking Data}
\label{perf_model}

For these tests, the implementations were tested on  matrices 
representing the transition probability functions of Markov chains based
on actual randomized algorithms. These matrices were then reduced to only $S_?$ states 
and subtracted from the identity matrix, as discussed in Section \ref{computing_reachability}.

JPF was used on two randomized algorithms to create this data.  The biased 
die algorithm simulates a biased dice roll by means of biased coin flips, and 
random quick sort is a randomized version of the quick sort algorithm. 
Both algorithms were coded in Java by Zhang, who also created the JPF 
extension that outputs transition probability matrices of Markov chains that correspond to the code being checked \cite{Xin}.

The JPF search strategies used were chosen to create the largest $S_?$ 
matrices relative to the size of the searched space.  JPF's built-in 
depth first search was used for random quick sort, and a strategy called 
probability first search that prioritizes high-probability transitions, 
also written by Zhang \cite{Xin}, was used for the biased die example.

\begin{figure}[!ht]
\centering
 \fbox{ 
\includegraphics[width=0.8\linewidth]{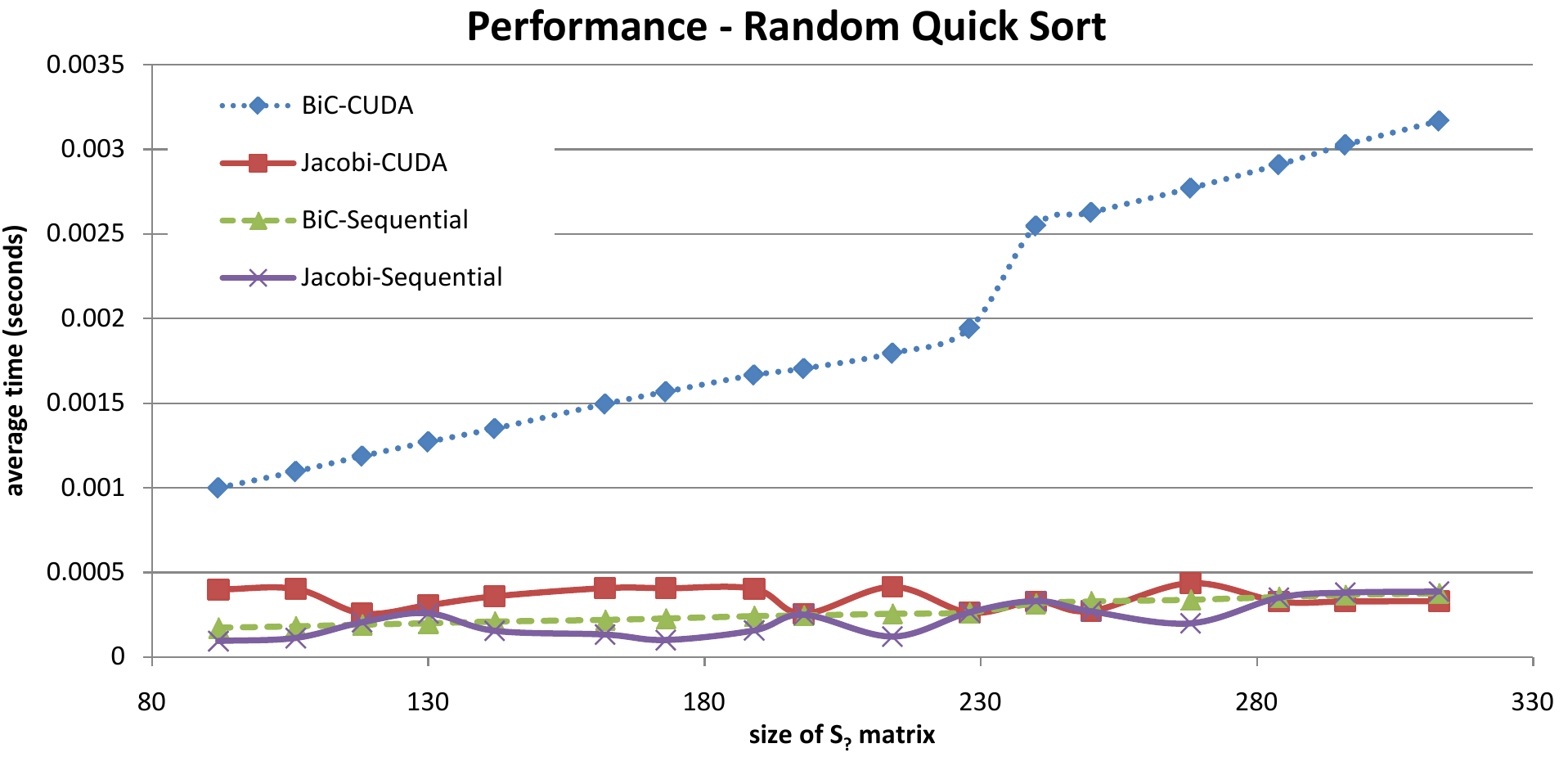}
}
\caption{Performance on matrices output by JPF, while checking the random quick sort algorithm.}
\label{RandomQS_JPF}
\end{figure}

\begin{figure}[!ht]
\centering
 \fbox{ 
\includegraphics[width=0.8\linewidth]{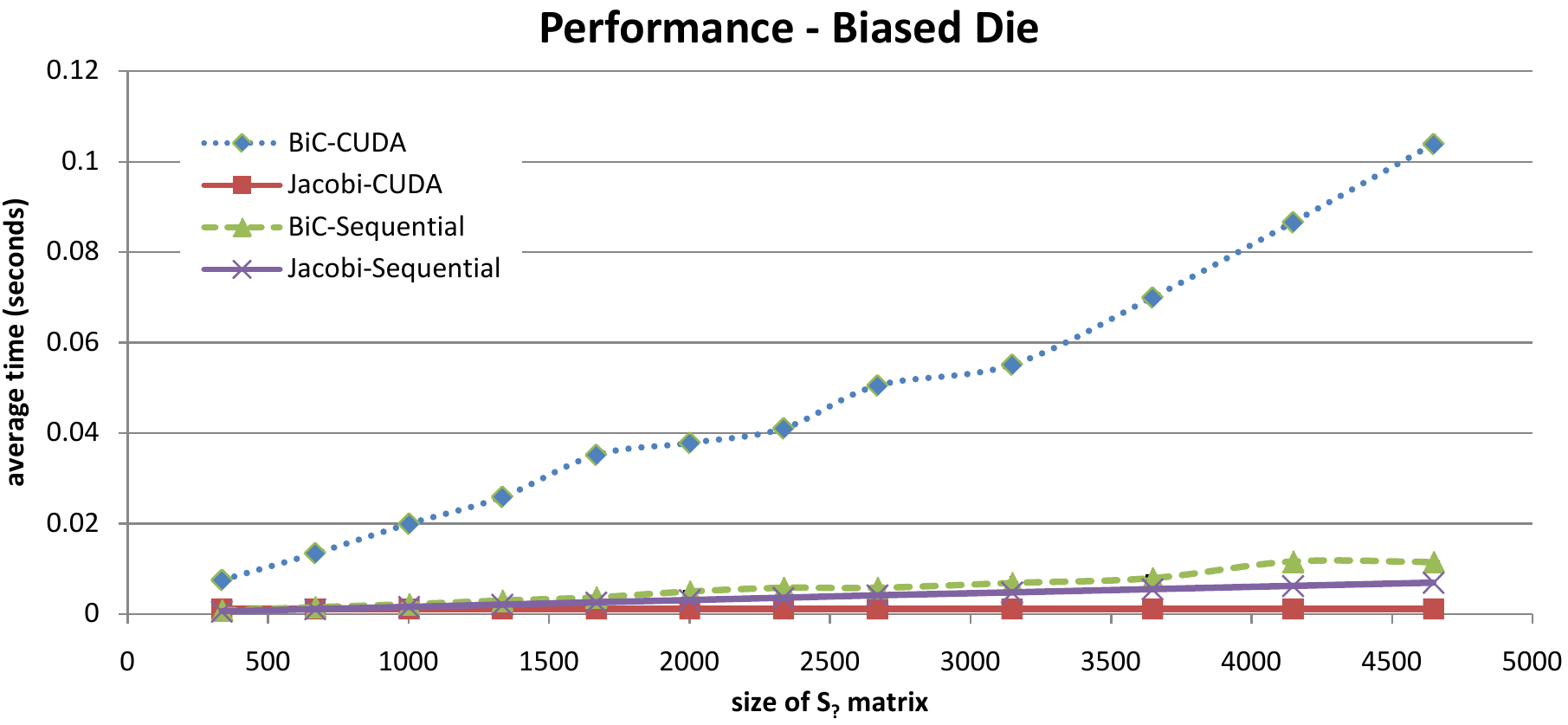}
}
\caption{Performance on matrices output by JPF, while checking the biased die algorithm.}
\label{BiasedDie_JPF}
\end{figure}

The results of the random quick sort tests are shown in 
Figure~\ref{RandomQS_JPF}, and the results of the biased die tests in 
Figure~\ref{BiasedDie_JPF}.  Error bars representing standard deviations 
of each data point are too small to be visible in the graphs.  The matrix 
sizes in the random quick sort tests are much smaller than those in the 
biased die tests because, while the matrices for random quick sort are 
initially the same size as or larger than those produced for biased die, 
much fewer states belong to the $S_?$ set.  Furthermore, note that the 
densities of these matrices decrease as their sizes increase.  The sizes 
and densities of the matrices used in these tests are shown in 
Table~\ref{table_densities}.

The relative performances of the sequential and CUDA implementations of 
the Jacobi method, and the sequential implementation of the BiCGStab method, 
are different for each algorithm.  The best performance on the random quick 
sort data was from the sequential implementation of the Jacobi method, 
while the best performance for biased die was from the CUDA implementation
of the Jacobi method. The CUDA implementation of the BiCGStab method performs 
worst in both cases.

\begin{table}
\small  
\centering
 \fbox{ 
  \begin{tabular}{ c c c | c c c}   
\multicolumn{6}{c}{ \textbf{ Sizes and Densities of $S_?$ Matrices } } \\ 
\hline
\multicolumn{3}{c |}{ Random Quick Sort} & \multicolumn{3}{c }{  Biased Die} \\
\hline

$n$	&	 $m$ 	&	density	&	$n$	&	 $m$ 	&	density	\\
92	&	211	&	0.025	&	667	&	1333	&	0.003	\\
118	&	263	&	0.019	&	1333	&	2665	&	0.001	\\
142	&	312	&	0.015	&	2000	&	3999	&	0.001	\\
173	&	379	&	0.013	&	2668	&	5335	&	0.001	\\
198	&	430	&	0.011	&	3647	&	7293	&	0.001	\\
228	&	491	&	0.009	&	4647	&	9293	&	0.000	\\
250	&	536	&	0.009	&	5647	&	11293	&	0.000	\\
284	&	606	&	0.008	&	6647	&	13293	&	0.000	\\
313	&	669	&	0.007	&	7647	&	15293	&	0.000	\\

  \end{tabular}
}
  \caption{Sizes and densities of the matrices produced by JPF for the two algorithms. $n$ is the matrix dimension, and $m$ is the number of non-zero entries.}
  \label{table_densities}
\end{table}

\subsection{Comparing Probabilistic Model Checking Data with Random Matrices}

For these tests, random matrices were generated with
the same densities and sizes as those produced by JPF (as in 
Table~\ref{table_densities}).  The JPF matrices are reduced transition probability
matrices, subtracted from the identity matrix. So, they have entries in the 
interval [-1, 1] with locations based on the structure of a Markov chain. 
In contrast, entries in the randomized matrices are randomly-located positive  
integers, as described in 
Section~\ref{perf_random}. Unlike in the JPF matrix tests, each trial 
uses a different matrix and vector.  However, the standard deviations 
of the times measured are still too small to be shown on the graphs.

\begin{figure}[!ht]
\centering
 \fbox{ 
\includegraphics[width=0.75\linewidth]{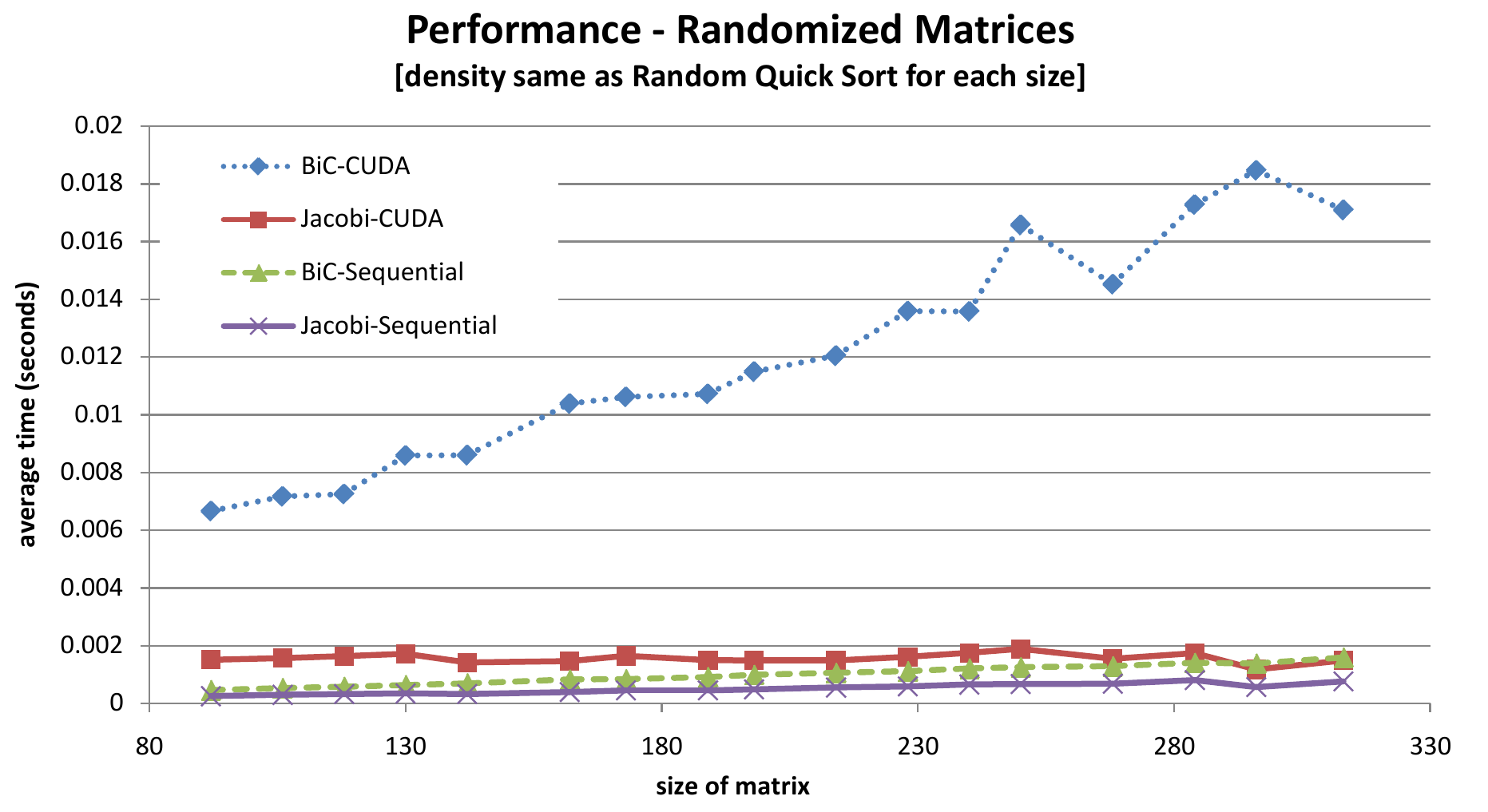}
}
\caption{Performance on randomized matrices, with the same sizes and densities as the matrices produced by JPF when checking the random quick sort algorithm. Averages times are based on 40 matrices.}
\label{RandomQS_randomized}
\end{figure}

\begin{figure}[!ht]
\centering
 \fbox{ 
\includegraphics[width=0.75\linewidth]{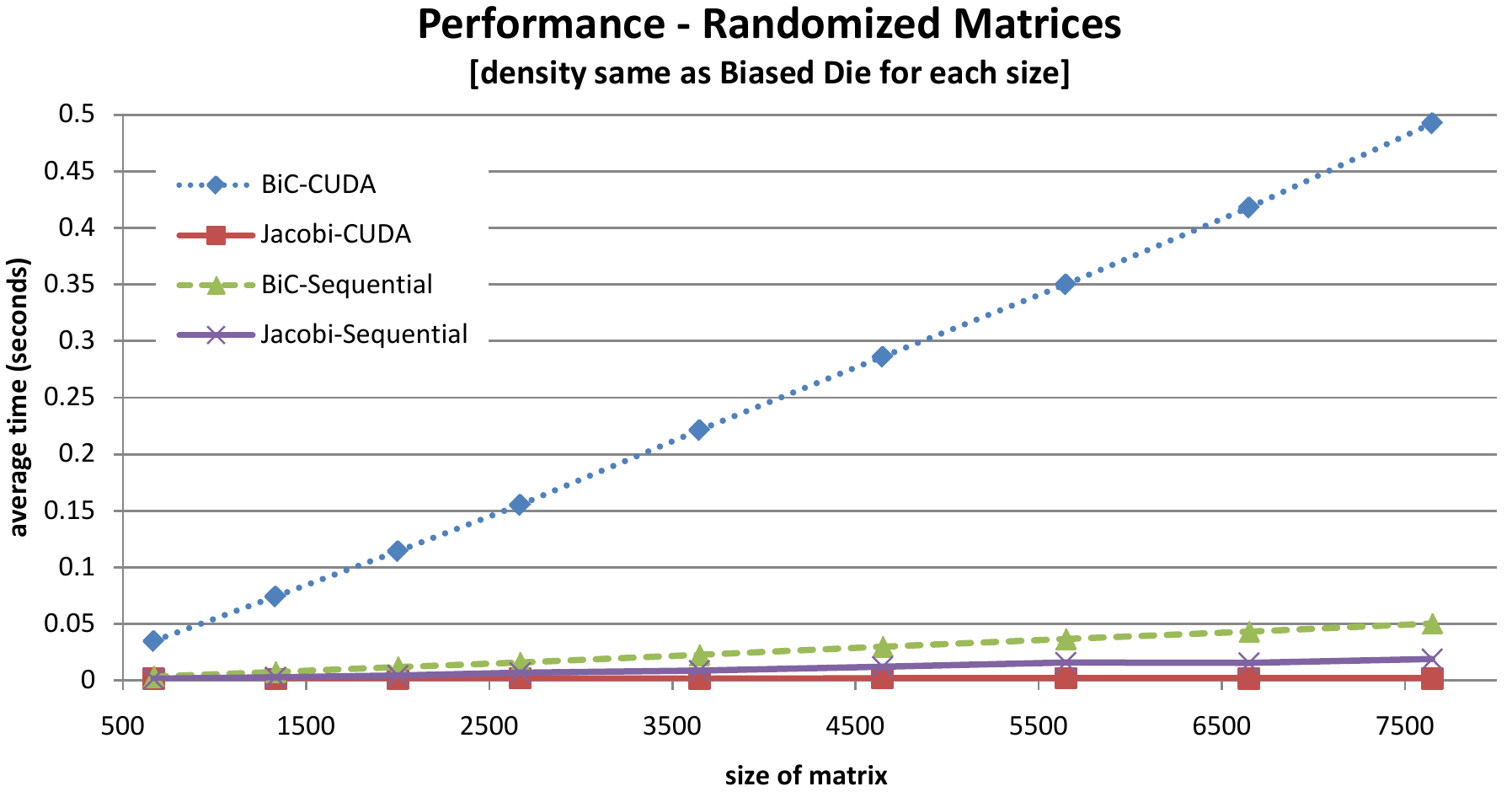}
}
\caption{Performance on randomized matrices, with the same sizes and densities as the matrices produced by JPF when checking the biased die algorithm.  Average times are based on 40 matrices.}
\label{BiasedDie_randomized}
\end{figure}

Performance results for these matrices are shown in 
Figure~\ref{RandomQS_randomized} and Figure~\ref{BiasedDie_randomized}. 
It is apparent that the ordering of the different implementations' 
performances is the same as it was for matrices of the same sizes and 
densities generated by JPF.  This suggests that size and density are the 
main determinants of which implementation performs best on probabilistic model 
checking data, and whether CUDA will be beneficial, rather than other 
properties unique to the matrices found in probabilistic model checking.

Generally, it appears that for the particular types of matrix encountered 
during probabilistic model checking, implementing the BiCGStab method in 
CUDA does not improve performance. CUDA does, however, improve the 
performance of the Jacobi method.

In \cite{Bosnacki}, the authors conjecture that the Jacobi method would 
be more suitable for probabilistic model checking using CUDA than 
Krylov subspace methods such as the BiCGStab method.  This seems to be 
true.  However, the results in Section~\ref{perf_random} suggest that this 
is due to the superior performance of the Jacobi method on sparse matrices 
in general, rather than BiCGStab's higher memory requirements as proposed 
in \cite{Bosnacki}.  For the probabilistic model checking data, the matrix 
density decreases as the size increases, so the conditions in which the 
CUDA BiCGStab implementation performs best (larger, denser matrices, as in 
Figure~\ref{Random_density10_all}) are not encountered.

\section{Related and Future Work}

Related work by Bosnacki et al. \cite{Bosnacki} tests a CUDA
implementation of the Jacobi method on data obtained by another 
probabilistic model checker, PRISM.  They used different probabilistic 
algorithms in their probabilistic model checker, and for each
algorithm their results show a benefit from GPU usage. In a later 
expansion of their research \cite{Bosnacki2}, they test a CUDA
implementation of the Jacobi method that uses a backward segmented 
scan, and one using two GPUs, which further improve performance. 
In their second paper they also compare the performance on 32- and 
64-bit systems, and find that for one of the three algorithms they 
model check, the CPU algorithm on the 64-bit system outperforms 
the CUDA implementation.  In another related paper, Barnat et al.\
\cite{Barnat} improve the performance of the maximal accepting 
predecessors (MAP) algorithm for LTL model checking by implementing 
it using CUDA.

Future work could include experiments on model checking data generated from the probabilistic algorithms used with the model checker in \cite{Bosnacki}, to allow closer comparison with that work. Another possibility is to implement the CUDA BiCGStab algorithm using multiple GPUs, so that different steps can be run in parallel.

In \cite{KPQ11}, Kwiatkowska et al.\ suggest to consider the 
strongly connected components of the underlying digraph of
the Markov chain in reverse topological order.  Since the
sizes and densities of the matrices corresponding to these strongly 
connected components may be quite different from the size and 
density of the matrix corresponding to the Markov chain,
we are interested to see whether this approach will favour
different implementations.

\medskip

\noindent
{\bf Acknowledgements}\ We would like to thank Stefan Edelkamp for providing us
their CUDA implementation of the Jacobi method so that we could
ensure that our implementation was similar. We also thank the referees for their
helpful feedback.

\bibliographystyle{eptcs}
\bibliography{submission}

\end{document}